\documentstyle[12pt,aasms4,flushrt]{article}

\begin{document}

\title{INTERMEDIATE MASS STARS: UPDATED MODELS}

\author{Inma Dominguez\altaffilmark{1},
Alessandro Chieffi\altaffilmark{2}, Marco Limongi\altaffilmark{3}, 
Oscar Straniero\altaffilmark{4}}

\affil{1. Universidad de Granada, Granada, Spain; inma@goliat.ugr.es}

\affil{2. Istituto di Astrofisica Spaziale - CNR, C.P. 67, 00044 Frascati, Italy; alessandro@altachiara.ias.fra.cnr.it}

\affil{3. Osservatorio Astronomico di Roma, Via Frascati 33, 00040 MontePorzio, Italy; marco@nemo.mporzio.astro.it}

\affil{4. Osservatorio Astronomico di Collurania, 64100, Teramo, Italy; straniero@astrte.te.astro.it}

\begin{abstract}

A  new  set  of  stellar  models in the mass range 1.2 to 9 $M_{\odot}$ is 
presented.  The adopted chemical compositions cover the typical 
galactic values, namely $0.0001 \le Z \le 0.02$ and $0.23 \le Y \le 0.28$. 
A comparison among the most recent compilations of similar stellar models is also discussed.
The main conclusion is that the differencies among the various evolutionary results are
still rather large. For example, we found that the H-burning evolutionary 
time may differ up to 20 \%. An even larger disagreement is found for the He-burning phase
(up to 40-50 \%). 
Since the connection between the various input physics and the numerical algorithms 
could amplify or counterbalance the effect of a single ingredient on the resulting stellar model, 
the origin of this discrepancies is not evident. 
However most of these discrepancies,
which are clearly found in the evolutionary tracks, are reduced on the isochrones. 
By means of our updated models  
we show that the ages inferred by the theory of stellar evolution 
is in excellent agreement with those obtained  by using other independent 
methods applied to the nearby Open Clusters. Finally, 
the theoretical initial/final mass relation is revised.
\end{abstract}


\section{INTRODUCTION}

The comprehension of the evolutionary status of the various stellar systems, from
the simplest stellar clusters up to the more complex galaxies, are mainly based 
on the comparison between theoretical stellar models and observational data.
It follows that any improvement and/or assumptions in the basic input physics (equation of state,
opacity, and the like) included in the computations
of the stellar models directly influences the interpretation of the observed data. 
Moreover, since the population synthesis requires the availability of
stellar models in a large range of masses and chemical compositions,
and since such an homogeneous database is still missing,
people involved in such kind of studies have been forced to collect models 
computed by different authors.

In this paper we present new evolutionary sequences for stellar masses ranging between 1.2 and 
9 $M_{\odot}$;
these models have been included in  our  database of stellar evolution which is  
available by anonymous ftp. This database also includes fully homogeneous models for low mass stars 
(Straniero, Chieffi \& Limongi 1997) and those for 
massive stars (Chieffi, Limongi \& Straniero, 1998, Limongi, Straniero \& Chieffi, 1998). 
All these models have  been  obtained by means of the FRANEC,   
an acronym of Frascati Raphson Newton Evolutionary Code (Chieffi \& Straniero, 1989,
and Chieffi, Limongi \& Straniero 1998).  The input physics (equation of state, opacity, neutrino losses 
and the like) we are adopting at the moment are discussed in Straniero, Chieffi \& Limongi (1997).

Several papers described sets of models of intermediate mass stars with various chemical compositions.
However, owing to the continues improvement of the input physics, the (re)computation of the same stellar models becomes 
necessary. 
Major contributions are from: Kippenhahn, Thomas \& Weigart 1965; Iben 1967a (and references therein);
Paczynski 1970a, 1970b, 1971; Trimble, Paczynski \& Zimmerman 1973; Alcock \& Paczynski 1978;    
Becker \& Iben 1979; Becker 1981; Vandenberg 1985; Maeder \& Meynet, 1987, 1988, 1989, 1991; 
Castellani, Chieffi \& Straniero 1990, 1992; Lattanzio, 1991; Stothers \& Chin 1990, 1991a, 1991b, 1992;
Vassiliadis \& Wood 1992; Alongi et al. 1991, 1992; Bressan et al. 1993; Schaller et al. 1992. 
Thus, our new models will be compared with the most recent compilations of similar evolutionary sequences.
Whenever possible, we analyze the origin of the resulting differencies. The main goal of this study is to constraint
the present level of uncertainty of the stellar evolution in the range of intermediate mass stars.
In such a way, we provide a basic tool to check the reliability of our understanding of the galactic
history as emerging from the study of population synthesis.

This is the plan of the paper: in the next section we revise the possible sources of uncertainties for 
H and He burning intermediate mass stellar models; in section 3 we 
describe our latest evolutionary
computations for intermediate mass stars, from the ZAMS up to the AGB;
in section 4 we discuss the comparisons among different evolutionary models; selected tests of the evolutionary
sequences are presented in section 5. Final remarks follow.

\section{INPUT PHYSICS AND CONVECTION}

If the theoretical investigation of low mass stars appears well anchored to 
the result of helioseismology (see Straniero, 
Chieffi \& Limongi, 1997) this is not the case  for  intermediate mass 
stars. It is commonly believed that the many uncertainties in the theory of turbulent convection  
still affects our understanding of the internal structure of these kind of stars. Owing to the lack of a conclusive 
test for the adequacy of the current theory of convection, the astrophysical literature presents a
variety of different approaches to the computation of stellar models.   
It has been early recognized that  
the mixing  of material in the core of a given star 
significantly  alters  its lifetime and in turn it could modify the 
age estimations of the various galactic components. The
instability against turbulent convection is classically handled 
by means of a thermodynamical criterion (namely the  Schwarzschild criterion 
for  a chemically homogeneous fluid). As it is well known, this criterion is 
based  on the evaluation of the expected gradient of temperature produced by 
the  radiative  transport of energy: when the required gradient is too high, 
the   radiative  flux  cannot  account  for  the  whole energy transport and 
hence convection is settled on. 

First of all let us emphasize that the correct evaluation of the size of an unstable 
region is primarily dependent on the accuracy of the input
physics. Any improvement of the 
stellar physics  (eos,  opacity,  cross  section  and  the  like) could imply a 
variation  of the estimated value of the temperature gradient and in turn it 
could  modify  the  location  of the borders of the convective regions, with 
sizable consequences on the computed stellar lifetime. 

A second question concerns the possibility that the convective motion is not 
drastically inhibited in a stable region located just outside the convective 
core.  In fact, although out of the Schwarzschild border a moving element of 
matter  is subject to a strong deceleration, it might be possible that a non 
zero velocity is maintained for a certain path. In 
such a case this {\it mechanical overshoot} might induce a mixing of material 
in  a  radiatively  stable  region  and  might also contribute to the energy 
transport. A large number of papers have been devoted to the inclusion of such
phenomenon in the computation of stellar models. Major contributions are from:
Shaviv \& Salpeter (1973); Maeder (1975); Cloutman \& Whitaker (1980); Bressan et al (1981);
Stothers \& Chin (1981, 1990); Matraka, Wassermann \& Weigert (1982); Xiong (1983, 1986); Doom (1982, 1985); 
Alongi et al. (1991); Maeder \& Meynet (1987, 1988, 1989, 1991); Shaller et al. (1992).
  
Unfortunately the available convection theory is still inadequate 
for  a  reliable  description  of  this  phenomenon  (see Renzini 1988 for a 
critical  discussion on this argument), so that the evaluation of the degree 
of  both  matter  and  energy  transport out of the unstable regions must be 
obtained   by   comparing  the  result  of  parameterized  models  with  the  
measurements  of some selected observable quantities (see e.g. Bressan et al. 1993). 
As for the {\it mixing length} (Chieffi, Straniero \& Salaris 1995),
the calibration of the 
free parameters used to describe the mixing in the overshooting region, is model
dependent. In fact, since the assumed input 
physics  affects  the  size of the unstable regions, the calibration of 
the  overshooting  depends  on these assumptions. For example, the larger is 
the opacity the larger is the  convective  region  and  in  turn  the  lower 
is  the required amount of overshooting. But the opacity coefficients are 
likely underestimated rather then overestimated. For this reason, if at the 
beginning  of  the  eighties   Becker \& Methews  (1983)  claimed a relatively 
strong  overshoot  in order to  reconcile the  theory with the observed  
distribution of stars in the young Globular 
Clusters of the Magellanic Clouds, the latest attempts to derive the size of 
the  convective  core  overshoot  for H-burning stars indicate that, if it's 
present,  it should be "mild" (see e.g. Sthothers \& Chin 1992; Castellani, 
Chieffi  \&  Straniero  1992; Schaller et al. 1992;
Bressan et al. 1993; Demarque et al. 1994, Mermilliod et al. 1994; 
Schr\"oder et al. 1997). In particular these studies generally  found that 
the  best reproduction of the various indicators of the convective core size 
is obtained with models including an overshoot roughly confined in between 0 
and  0.3 (in units of pressure scale height and measured from above the stability
border defined by means of the Schwarzschild criterion).

The situation is still more controversial for the central He-burning phase. In such a case,
when He is converted into C within the core, the opacity increases and, in turn, the convective core
size must increase (Schwarzschild 1970; Paczynski 1970b; Castellani et al. 1971a).
As the He burning proceeds,
a minimum in the radiative gradient settles on, so that mixing occurs in two separated regions: an internal
one, which is fully convective, and an external one, in which the resulting mixture of C and He is just that
needed to 
allow the convective neutrality (Castellani et al, 1971b; see also Iben 1986 and references therein). 
Such phenomenon was called {\it He burning semiconvection}.
Close to the He exhaustion (namely when the central He becomes lower then approximately 0.1), some instabilities
at the border of the convective core appears in stellar model computations (Castellani et al. 1985 a,b; Iben 1986).
These instabilities were called {\it breathing pulses} (BP). As a consequence of both semiconvection and BPs, a larger
amount of fuel is available for the central He-burning. 
There is some debate concerning the actual occurrence of the BPs in real stars
(Renzini \& Fusi Pecci, 1988; Caputo et al., 1989). In any case,
the inclusion of these phenomena might affect some important results of stellar evolution:
the estimated central He-burning and AGB evolutionary times, the final amount of
C and O in the core, the final WD mass.
Note that the efficiency of both semiconvection and BPs
also depend on the adopted input physics. For example, as firstly discussed by Iben (1972),
the use of different prescriptions for the  
$^{12}C(\alpha,\gamma)^{16}O$  reaction rate alters the duration of the final part of the
He-burning phase during which the BPs occur. Then, a larger value of this rate will 
enhance the effects of the BPs.

In summary, when comparing the evolutionary results obtained by different authors, the connection between     
the assumed mixing scheme and the adopted input physics must be taken into account.
This will be done in section 4.

\section{THE NEW MODELS}

Models  from  1.2  to  9  $M_{\odot}$ and metallicity ranging between $10^{-4}$ and 
$2 {\times} 10^{-2}$ have been computed from the ZAMS (Zero Age Main Sequence) up 
to  the  end of the E-AGB  (Early  Asymptotic Giant Branch) phase. The evolutionary tracks
in the HR diagram are reported in figures 1, 2, 3 and 4. The runs of the central temperature 
versus the central density are shown in figures 5, 6, 7 and 8. Some examples of
the evolution of the fully convective regions
are illustrated in figures 9, 10, 11 and 12.

\placefigure{f1int}

\placefigure{f2int}

\placefigure{f3int}

\placefigure{f4int}

\placefigure{f5int}

\placefigure{f6int}

\placefigure{f7int}

\placefigure{f8int}

In tables 1 to 4 we 
have  reported  the  fundamental properties of the evolutionary sequences, 
namely from column 1 to 10: the total mass, the central  H-burning lifetime (in Myr), 
the  maximum  size  (in  solar  mass unit) of the convective core during  
central  H-burning,  the surface He mass fraction after the first dredge-up, 
the  tip  luminosity of the first RGB (red giant branch), the He 
core  mass at  the  beginning of the He-burning, the central He-burning lifetime (in Myr), 
the  He  core mass (in solar unit) at the end of the He burning, the surface 
He  mass  fraction  after  the  second dredge-up and the He core mass (in solar 
unit) at the beginning of the TP-AGB (thermally pulsing asymptotic giant branch) 
phase.

\begin{table}
\dummytable\label{table1}
\end{table}
\placetable{table1}

\begin{table}
\dummytable\label{table2}
\end{table}
\placetable{table2}

\begin{table}
\dummytable\label{table3}
\end{table}
\placetable{table3}

\begin{table}
\dummytable\label{table4}
\end{table}
\placetable{table4}

In the following part of this section we briefly summarize the main features of the computed sequences of models, 
revising the dependence of the various evolutionary phases on the stellar mass and on the chemical composition.
As already recalled in the introduction of this paper, the evolutionary history of an intermediate  mass star, 
crossing the HR diagrams from the main sequence up to the AGB,
is well known. For a more accurate description of the various evolutionary phases and an exhaustive
list of references we remind the reader to the review paper by Iben (1991). 

\subsection{THE CENTRAL H-BURNING}

All the  sequences  of  models  having  mass  larger  than 1.2 $M_{\odot}$
develop a convective core during the central H burning
independently  on the initial chemical composition.
As a consequence, at variance with low mass stars for 
which  the  H  burning  occurs  in a radiative environment, the evolutionary 
tracks  evolve off the ZAMS towards lower temperature and larger luminosity. As
the H is converted into He in the central region of the star, the opacity decreases and 
the convective core recedes (in mass). The convective instability in
the core is retained until the H mass fraction 
is  reduced  down  to  about 0.1. Then an overall contraction occurs and the 
star  rapidly  move  towards the radiative  main  sequence.  A  maximum  in  
luminosity is reached at the time of the H exhaustion. 

\placefigure{f9int}

\placefigure{f10int}

\placefigure{f11int}

\placefigure{f12int}

One interesting quantity characterizing the H-burning phase is the maximum extension of
the convective core. As already recalled, this maximum is attained just after the ZAMS
(see figures 9, 10, 11 and 12). In the last 
30 years the computed values for this quantity have been systematically increased, mainly due to the 
increasing values of the adopted radiative opacity coefficients. Our present results are listed in 
column 3 of tables 1 to 4. Note that the size of the convective core decreases monotonically as the
initial He increases while it initially increases as the metallicity decreases (up to Z=0.001) and then it 
decreases at smaller metallicities.
The corresponding H-burning evolutionary times are reported in column 2.

\subsection{THE H-BURNING SHELL}

When  the  H-burning shell settles on, the convective envelope penetrates
more  than 80 \% in mass of the star, bringing to the surface the products of 
the  H  burning.  As firstly pointed out by Iben (1964, 1967b),
the  main  consequences  of  this first dredge-up are: the 
increase  of  the surface abundances of $^4$He, $^3$He and $^{14}$N and a decrease 
of  those  of  $^{12}$C  and  $^{16}$O. The modification of the 
surface  composition  is stronger in low mass stars due to the minor size of 
the  envelope.  The surface amount of He resulting  after the first dredge-up in our models is 
listed in column 4 of tables 1 to 4.
The subsequent  evolution up to the onset of the He burning 
phase  is  mainly  characterized  by  the equation of state governing the He 
core.  In  the  strong  degenerate  regime  a  quite  large  He core mass is 
necessary  to  ignite  He (namely about 0.5 $M_{\odot}$, but it depends on 
the chemical composition). This is the case of a low mass star (i.e. $M \le 1.5 
M_{\odot}$).  For  more massive stars the degree of degeneracy in the core is 
reduced  and  in  turn  the  He  ignition  is  more rapidly attained. In the 
asymptotic  limit of non degenerate matter the minimum mass needed to ignite 
He  is about 0.35 $M_{\odot}$. For this reason the luminosity of the RGB tip and 
the  He core mass attained at the He ignition (columns 5 and 6 of tables 1 to 
4)  decrease  when  the  total  mass increases from 1.5 up to 2.5 $M_{\odot}$. 
When the stellar mass is larger than 2.5-3 $M_{\odot}$, the off main 
sequence  evolution is not further controlled by the growth of the He core. In 
such  a  case, owing to the internal mixing occurring during the main sequence, 
the  H-burning shell at the beginning of the RGB settles well outside the 
minimum  mass  needed to ignite He. Hence, the RGB tip and the mass of the He 
core  at  the  He ignition rise as the total mass increases.

\placefigure{f13int}

The  
minimum resulting by the combination of these two behaviors marks the so called RGB 
phase transition (Iben 1967c). Such an occurrence is 
illustrated  in  figure 13. According to the classical results, 
we found that the minimum core mass is attained for a 
total mass of 2.3-2.5 $M_{\odot}$, value slightly increasing with the metallicity. 
Note that the almost constant minimum core mass at very different metallicities is the consequence
of the opposite influence of Z and Y on this quantity.
In fact, as pointed out  by Sweigart, Greggio \&  Renzini (1990), the transition mass increases  
as the metallicity increases and decreases as the He increases.

\subsection{THE CENTRAL HE-BURNING}

As  it  is  well  known,  the  larger  the  core mass at the He ignition the 
brighter is the star during the central Helium burning phase and the shorter is the central He 
burning lifetime. Hence a maximum in such a lifetime 
is  expected, corresponding to the minimum He core mass occurring at the RGB phase 
transition.  In  figure 14 we have reported this lifetime for our models  
with Z=0.02 and Y=0.28. They are listed in column 7 of tables 1 to 4 for the full set of models. 

\placefigure{f14int}

During central He burning the evolutionary tracks move towards the blue 
part  of the HR diagrams on a Kelvin-Helmotz time scale and then moves back to 
the red giant branch when the central He vanishes. The extension of this 
loop depends on both the stellar mass and the chemical composition (see Alcock \& Paczynski, 1978,
and references therein). The larger 
the  mass  the hotter the left border of 
the  loop, but an interesting exception is worth to be noted (Castellani, Chieffi \& Straniero, 1990).
At the lower metallicities
(Z=0.0001, 0.001 and 0.006), the more  massive  stars  ignite  He  before they can reach the RGB,
thus skipping the first  dredge-up.  It  occurs  for  stellar masses $M\ge2.7 M_{\odot}$,
$M\ge4M_{\odot}$ and $M\ge5M_{\odot}$, at  Z=0.0001, 0.001 and 0.006 respectively.
Since the larger is the surface He amount the lower is the opacity, the fate of the first dredge-up affects 
the blue loop extension: those models in which the first dredge up do not occur 
have a narrower blue loop. Such an occurrence is clearly shown in figure 2, which reports the HR
diagrams for Z=0.006: note that the He burning evolutionary track of the $4 M_{\odot}$ sequence have a
significantly larger blue loop than the $5 M_{\odot}$ one.

\subsection{THE DOUBLE SHELL PHASE}

During  the whole central He burning phase the H-burning shell moves outward 
so that the longer is the central He burning lifetime the 
greater will be the increment of the He core mass (see column 8 in tables 1 to 
4).  For this reason, the minimum in the $M_{He}$-initial mass relation, which 
is  evident  at the beginning of the central He burning, is smoothed away at 
the  end  of the E-AGB phase (see figures 15). As a consequence all 
the  stars  with $M\le3$ $M_{\odot}$ starting the thermally pulsing AGB phase
have a rather similar He core mass, namely $0.55 \pm 0.05 M_{\odot}$ depending
on the metallicity (column 10 of tables 1 to 4). Such a 
value provides us a lower limit to the expected mass of a CO WD. 

\placefigure{f15int}

For a sufficiently high initial stellar mass a second dredge-up occurs 
during the early AGB (Kippenhahn et al. 1965, Paczynski 1970, Becker \& Iben 1979).
In such a case, the convective envelope
penetrates the H discontinuity located just below the H-burning
shell so that the resulting He core mass is lowered with 
respect to the value attained at the end of
the central He-burning (column 10 and 8 of tables 1 to 4, 
respectively). The second dredge-up takes place only if the H-burning shell extinguishes. 
This is not the case for a low mass stars, in which the expansion induced  by 
the He-burning shell did not induce a sufficient cooling of the H-burning one.
We found that the minimum mass for the occurrence of the second dredge-up is 4 $M_{\odot}$ for
the three lower metallicities and a bit larger
(about 5 $M_{\odot}$) in the case of $Z=0.02$. The surface He mass fraction after the second dredge-up is
listed in column 9 of tables 1 to 4.

\subsection{THE FINAL MASSES AND $M_{UP}$}

The white dwarf (WD) masses resulting from the evolution of low and intermediate mass stars are very 
important quantities for the purpose of the study of population synthesis,
Planetary Nebula, Novae, Super novae, and the like (see e.g. Iben, 1991).

In figure 16 we compare our theoretical final masses 
with the relation reported by Weidemann (1987) and the updated one by Herwig (1995).
Squares represent the He core masses at the end of the  E-AGB, while arrows show the growth of the 
He core masses during the TP-AGB phase, as derived by using our thermally pulsing models
(Straniero et al. 1996).
The number reported at the top of each arrow is the
number of thermal pulses computed up to the end of the AGB phase. 
It was determined according to the mass loss rate prescriptions of
Groenewegen \& de Jong (1994).

\placefigure{f16int}

Note that the Weidemann predictions (as well as the recent improvements incorporated by Herwig) 
are based on a semi empirical approach. 
They make use  
of various methods to 
evaluate the WD masses in nearby Open Clusters, whose initial mass is derived from the turnoff age.
Our final masses are instead the result of a pure 
theoretical calculation and, hence, they are mainly dependent on the adopted input physics.
Such a difference should be well kept in mind when
comparing our results with those of Weidemann (or Herwig), as we do in
figure 16.  
Despite the two different approaches, there is an acceptable agreement between our theoretical final masses and 
those of Weidemann (Herwig).
In the next section we will discuss our final core masses in comparison with the ones obtained by other
authors by means of alternative stellar evolutionary codes.

Stars with higher masses ignite the Carbon before the
onset of the thermally pulsing phase 
(Paczynski 1971; Alcock \& Paczynski 1978; Becker \& Iben 1979,1980;
Castellani, Chieffi \& Straniero 1991; Bressan et al. 1993;
Garc\'ia-Berro, Ritossa, Iben 1997). In tables 1 to 4 we have
distinguished among the models which experience an off center
Carbon ignition and those with a central Carbon ignition. We found 
that $M_{up}$, i.e. the maximum mass for which the concurrent 
action of the pressure of a 
strong degenerate electron component and the neutrino energy loss 
in the core prevent the onset of the C-burning, ranges between 6.5 and
8 $M_{\odot}$, the lower and the 
larger values being obtained for Z=0.0001 and 0.02, respectively

\section{THE PRESENT LEVEL OF UNCERTAINTY}

Owing to the large amount of numerical algorithms and  physical ingredients commonly used in the 
computation of stellar models,
the evaluation of their reliability is not trivial. A first idea of the possible sources of uncertainties
can be obtained
by comparing the evolutionary sequences obtained by different authors by means of different   
evolutionary codes and/or input physics. This might be also useful to evaluate the correctness of
merging different sets of stellar models. As already recalled, there exist a rather large
number of papers which present set of models for intermediate mass stars. 
In the following we will compare our results 
with the most recent and widely adopted collections of these stellar models. 

\subsection{{\it OLD} AND {\it NEW} PHYSICS}

Let us firstly compare the present computations to
the ones of Castellani, Chieffi \& Straniero (1990 and 1992), 
which were obtained by means of almost
the same evolutionary code, but by adopting an "old" physics. Such a comparison 
will provide us with an evaluation 
of the importance of the most recent (last decade) improvements of the input physics.
In figure 17 we have compared the evolutionary tracks for Z=0.02.
The two sets of models appear rather similar
except for some (important) details.
Concerning the main sequence, the new tracks are slightly brighter and  
the convective path (i.e. from the ZAMS up to the beginning of the overall contraction)
is longer.

\placefigure{f17int}

The new H-burning lifetimes are generally lower (${\sim}5$ \%) than the old ones, but this   
is mainly due to the slightly lower amount of He used in our old computations (namely Y=0.27). 
Concerning the He burning, the most striking difference is the extension of the blue loop in 
the more massive sequences, the new ones being significantly wider.
The He burning lifetime is substantially unchanged.

\subsection{DIFFERENT EVOLUTIONARY CODES}

The second step in the evaluation of the reliability of the evolutionary sequences will be the comparison with 
the most recent and widely adopted compilations of stellar models. Let us distinguish between models with
and without mechanical convective core overshoot.

The most recent sets of intermediate mass stellar models without
overshoot have been published by Lattanzio (1991, L91) and Vassiliadis and Wood (1993, VW93). Despite the differences
in the chemical composition and in the input physics there is a good agreement between our H-burning models 
and the ones found in the two papers cited above. For example, by interpolating on the grid published by Lattanzio 
we derive for a $2.5 M_{\odot}$ (Z=0.02 and Y=0.28) an H-burning lifetime of 512 Myr to be compared 
with our result, namely 505 Myr. For the same stellar mass, but Y=0.25, Vassiliadis and Wood found 619 Myr,  
Since in this range of mass and metallicity we found that by increasing the original helium of 
${\delta}Y=0.1$ the corresponding $t_H$ must be reduced of about 25 Myr, the quoted value correspond to about 
545 Myr at Y=0.28. Similar differences are found for other masses and other chemical compositions.
Concerning the He burning models the situation is more complicated.  
The He burning lifetime is strongly dependent on both the assumed scheme for convection and the 
$^{12}C(\alpha,\gamma)^{16}O$ reaction rate. As in VW93, we allow semiconvection and suppress     
breathing pulses, whereas they are both allowed in the Lattanzio's computation. On the other hand, we use  
the $^{12}C(\alpha,\gamma)^{16}O$ reaction rate of Caughlan et al. (1985) which is about three time larger
than the rates adopted by Lattanzio (1991) and Vassiliadis and Wood (1993). We recall that the larger this reaction
rate the larger the He-burning lifetime.
Bearing in mind these differences in the input physics,  
for the $5 M_{\odot}$ with Z=0.02 (0.016 in VW93) one find
20.8, 23.5 and 30.6 in the present paper, VW93 and L91, respectively. For lower masses and/or metallicity 
the differences are similar. For example,  
for $M=1.5 M_{\odot}$ and Z=0.001 we found $t_{He}=97.4$ Myr to be compared to 122.2 Myr (VW93) 
and  118.3 (L91). 
In summary, our H-burning lifetimes appear in good agreement with those obtained in other studies,
being the differences in the evolutionary time scales
always lower than $10\%$. Note that similar differences were found with respect 
to our old computations. 
On the contrary, the present uncertainty in the theoretical evaluation 
of the He-burning lifetime is 
definitely larger. Differences up to $30\%$ are found in $t_{He}$. 
In principle they should be primarily attributed to the uncertainties in 
the convective algorithm and/or in the major He burning reaction rates. 
In practice, due to the connection between input physics and mixing efficiency, 
it is rather complicated to disclose the origin of such differences.

Recent models including a moderate amount of convective core overshoot  
have been published in a series of papers by the Padua group (see Bressan et al., 1993,
and references therein; in the following B93) and by the Geneve one (Schaller et al. 1992 and 
reference therein; in the following  S92). We recall that the B93 models
were obtained by extending, in practice, the  
mixed central region of an H-burning or He burning intermediate mass star by approximately 0.25 $H_p$ 
over the unstable zone, while Schaller et al. assume 0.2 $H_p$. 

\placefigure{f18int}

Concerning the H-burning,
when  the convective  core  overshoot is taken into account, a larger amount of 
fuel  is  available  in  the  burning region of the star. However, since the 
larger is the mixed region the brighter is the star, this additional fuel is 
more  rapidly  burned,  so  that  the effect of the core overshoot on the H-
burning lifetime is partially counterbalanced. In figure 18 we compare our H-burning lifetimes to
those resulting from the B93 and S92 models. As expected, overshoot models are generally older than the 
corresponding classical ones. Note, however, that despite the similar amount of overshoot 
assumed by Bressan et al. and Schaller et al., the differences  between these two sets
of models are comparable to the ones found with respect to our (no overshoot) models. 

Another important consequence of the convective core overshoot during the central H-burning 
is the reduction of the mass at which the RGB transition occurs.
Because the RGB evolution is faster if the star has a non degenerate He core, 
the number of stars lying on 
the RGB of galactic open clusters might be used, in principle, to derived the value of the 
transition mass and, in turn,
to discriminate in between models with and without 
core  overshoot (see e.g. Mermilliod  et al., 1994 and references therein).
The comparison  between our models and 
those  of Padua,  shows  that  the  
difference  is presently quite small. For example, at Z=0.02 we found
a {\it transition} mass of 2.4 $M_{\odot}$ while Bressan et al. (1993)
found 2.2 $M_{\odot}$. Thus, minor differences are aspected in the synthetic
RGB populations. Note that a similar comparison cannot be made with the Schaller et al. models 
because their set is not spaced enough in mass.

\placefigure{f19int}

At  variance  with the H-burning phase, the inclusion of a moderate overshoot 
in  computing  He  burning  models do not significantly alter the 
resulting  He-burning  lifetime. In fact, 
if a moderate  overshoot  is  taken  into  account, the semiconvective layer is 
hidden  by  this  extra mixing, but 
the total amount of fuel (He) available  for  the  central  nuclear burning should be 
practically the same than 
that found in models  without  core  overshoot but including semiconvection.  
Our He-burning evolutionary times and those obtained by Bressan et al. and Schaller et al. 
are compared in figure 19. Note that B93 adopt the rather low Caughlan \& Fowler (1988) rate for the   
$^{12}C(\alpha,\gamma)^{16}O$ reaction, whereas S92 use our preferred rate (i.e. 
Caughlan et al., 1985). The differences in the stellar lifetimes are larger than 
those found in the case of the H-burning phase (up to 60\%). Also in this case, 
the origin of the disagreement is not easily recognized. 

Let us conclude by noting that the typical
differences which we found when comparing our models (no overshoot) with those by L91, VW93, B93 and S92   
are of the same order of
magnitude of those found in the comparisons   
between the two set of models with convective core overshoot.
In other words, in the range of 
intermediate mass stellar models, the current uncertainty due
to a possible not negligible occurrence of a convective core overshoot appears less severe, 
or of the same order of magnitude, than those induced 
by other input physics.

\subsection{THE RELIABILITY OF THE THEORETICAL CORE MASSES}

In the previous section we have compared the final masses obtained by evolving our models up 
to the end of the AGB to 
the semiempirical initial/final mass relation (Weidemann, 1987).
These quantities depend on the core mass attained at the beginning of the thermally pulsing AGB phase and 
on the AGB mass loss rate (see e.g. Iben \& Renzini 1983).
For the more massive stars also the efficiency of the second 
dredge-up should be taken into account (Paczynski 1971, Becker \& Iben 1979,1980)
In the following we compare our evolutionary core masses at the first TP with the 
ones obtained by other authors. Let us recall that 
the larger the duration of the He burning phase the larger is the time available 
for the shell H-burning to advance in mass.
Hence, the large uncertainty on the current estimation of the stellar lifetime 
(as illustrated in section 4.2) 
might  affects the theoretical previsions of the final masses. 
Concerning the convective algorithm, the lowest lifetime, and in turn the smallest He core mass, 
is obtained when semiconvection, BPs and overshooting are neglected, while an 
approximate doubling of the He-burning lifetime is
found when, as in our models, only the semiconvection is taken into account. However 
by comparing  the RGB, HB and AGB theoretical lifetime ratios to the observed stellar
number ratios of well studied galactic Globular Clusters, it is possible to discriminate
among the various mixing hypothesis (see e.g. Renzini \&
Fusi Pecci 1988). In such a way, there is a support to the classical
semiconvection scheme (no BPs), but a moderate overshooting, which could mimic the effect of  
semiconvection, cannot be ruled out.
The current uncertainty in the resonant contribution to the 
$^{12}C(\alpha,\gamma)^{16}O$
reaction rate might also change the estimated He burning lifetime (Iben 1967). 
In such a case, by varying the astrophysical factor in the range of value compatible
with the available measurements for this reaction rate 
(see Buchmann, 1996 and 1997), we have obtained a variation of the
He burning lifetime of about 5-10 \%.

\placefigure{f20int}

In figure 20 we compare our core masses at the end of the E-AGB to those computed by Lattanzio (1991) 
and Bressan et al (1993). 
In spite of the rather large discrepancies found in the He-burning evolutionary time scales,
there is a good agreement between the core masses obtained by the 
different authors. Only the core masses of the more massive models by 
Bressan et al. are rather larger 
than ours (i. e. for $M \ge 5 M_{\odot}$). 
From table 5 of B93 we see that the maximum core masses
attained before the onset of the second dredge-up in the 5 and 7 $M_{\odot}$ sequences 
are slightly lower than in our models. Thus their larger core masses at 
the first TP should be a consequence of a 
less efficient second dredge-up. This is also confirmed by the smaller changes 
induced by the dredge-up on the surface composition. 
Note that B93 even include 0.7 $H_p$ of undershooting in their computations.

\section{TEST OF THE EVOLUTIONARY SEQUENCES}

Although  a  detailed  comparison  of  our  stellar models with the observed 
properties  of  different  stellar systems is well beyond the purpose of the 
present  paper,  let  us  discuss  two interesting tests of the evolutionary 
sequences  recently  proposed  by different authors, which allow us to check 
the  reliability  of  the  current  theoretical scenario. To do that we have 
computed selected isochrones and synthetic diagrams based on the present set 
of stellar models. 

\subsection{THE PLEIADES AND THE BROWN DWARF TEST}

The certain identification of brown dwarfs would provide important informations on 
the star formation rate of very low mass stars and 
contribute to shed some light on the dark matter problem. For this reason, in the last 
few  years, the search of these objects in nearby stellar clusters has been 
intensified.  Brown  dwarfs  are  (quasi) stars for which H burning does not 
occur  or,  at  least,  it  does  not  reach  the  full  equilibrium. How to 
certificate  such  an  occurrence?  A  low  mass  stars approaching the main 
sequence  is  fully convective, so that the products of the internal nuclear 
burning  should  appear  at  the  surface.  Thus, the best indicators of the 
occurrence of the internal nuclear burning are the secondary elements of the 
pp-chain.  In this context, Li is a good tracer, since it is a very volatile 
element  and  it  is  easily observable in faint objects. Stars with 
$M\le1 M_{\odot}$
deplete Li even in pre-main sequence. But in very low mass stars, 
when  the  internal  temperature  never exceeds $2-3 {\times} 10^6$ K, the Li remains 
unburnt.  Therefore,  the presence of observable Li lines in stellar spectra 
is  a  certain  identification  of  a brown dwarf (Rebolo, Martin \& Magazzu 
1992). 
In  turn  the  reappearance  of  Li  in  the  lower main sequence of stellar 
clusters  provides  an  interesting  test  for the age estimated by means of 
stellar  models.  In  fact  the  upper  luminosity  for which Li is measured 
depends  on the age of the cluster: the larger is the age the fainter is the 
Li  cutoff.  Basri  et  al.  (1996) by means of accurate infrared photometry 
coupled  to high resolution spectroscopy of brown dwarfs candidates in the 
Pleiades,  have  been  able to identify the Li cutoff. By using this Li 
test  they  found  an age of $\sim$ 115 Myr. They concludes that this value is 
definitely larger than the age estimated by comparing the turnoff luminosity 
with  that predicted by canonical stellar models.

\placefigure{f21int}

In  figure 21 we report the isochrones fitting obtained by 
using  our  present  stellar models (no overshoot) and that 
derived by Bertelli et al. (1994) by using 
the B93 models. Accordingly to the 
Hypparcos  parallaxes,  we adopt a  true  distance modulus of 5.33. A  reddening of 
0.04 has been assumed. In both cases the isochrones are computed for Z=0.02 
and  Y=0.28.  Concerning  the  canonical  models,  by excluding the isolated 
bright  star  at V=2.87, or $M_V=-2.58$, (a blue straggler?), one can get an  
age  of  at  least  120  Myr  and  not  exceeding 140 Myr, i.e. a value in very good 
agreement  with  the brown dwarf test. It is worth noting that owing to the 
few  amount  of stars in the turnoff region, a precise age cannot be derived 
by means of the isochrone fitting. However, even taking into account such an 
uncertainty,  we  can  exclude  ages  lower than 100 Myr. Similarly with the 
Bertelli  et al. (1994) isochrones one may get an age in between 150 and 200 
Gyr, which is a bit larger then that implied by the brown dwarf Li cutoff. Obviously, 
also in this case the uncertainty due to the few statistical significance of 
the  number  of  turnoff stars might be claimed, so that we can't definitely 
rule out the presence of a moderate overshoot. Let us finally note that this 
canonical estimation of the Pleiades age, based on the new distance, removes 
the  old  controversy  of  the lack of an evident lower main sequence turnon 
formed  by  those stars approaching the ZAMS (Herbig, 1962; Stauffer, 1984). 
In  fact the corresponding lower limit claimed by Stauffer (i.e. 100 Myr) is 
well in agreement with the present determination.

\subsection{THE WHITE DWARFS LUMINOSITY FUNCTION}

An intermediate mass star ($M \sim 5-6 M_{\odot}$) ends its life as a CO  
white  dwarf.  Thus,  if  this star is a member of an old stellar 
system  (say  1 Gyr or older), it spent most of its life as a WD so that  
its  cooling  time  might  be used as an age indicator for the stellar 
system.  A  search  for  the  cutoff  of the WD luminosity function has been 
recently  performed  by  Von Hippel et al. (1995) by means of the HST with 
the Wide Field Planetary Camera 2.  They  were  able  to  identify  this 
cutoff in two old open cluster,  namely  NGC  2420  and  NGC  2477.  Then,  
by means of the theoretical cooling sequences  computed  by  Wood  (1994),  
they estimated the ages of these two clusters  and  concluded  that  they  
are in contrast with all the available stellar  models.   Owing  to the 
existence of an accurate CCD photometry  only for one  of these two cluster, 
namely NGC 2420 (Anthony-Twarog et al 1990), we  will  focus  our attention 
on this one. From the paper by Von Hippel et al. we  derive an age based on 
the WD luminosity function cutoff of 1.5-1.6 Gyr.

\placefigure{f22int}

In  figure 22  we  show  the isochrones fitting to the 
 Color Magnitude  diagram of NGC 2420. According to Anthony-Twarog et al. 
 (1990), we  have assumed a metallicity of Z=0.008, which correspond to 
 about [M/H]=-0.4,   and  an  E(B-V)=0.05.  Then  the  distance  modulus  
 was  derived  by  reproducing  the clump of the He burning stars which is a feature
 almost independent on the age (see Castellani, Chieffi \& Straniero, 1992).
 The resulting age is of the order of 1.6 ($\pm$ 0.2) Gyr, value in very 
 good agreement  with  the  age  derived  from the WD cooling sequence. A 
 slightly larger  value would be obtained by adopting isochrones including a moderate 
 amount of convective  core  overshoot.  For example, Carraro \& Chiosi 
 (1994) found 2.1 Gyr while Friel (1995) reported 2.8 Gyr.  However, as  
 already  noted  by Demarque et al. (1994), the canonical isochrones  cannot 
 account for the distribution of stars near the turnoff of NGC 2420. In 
 particular the path of the isochrones just before the overall contraction  
 appears  shorter than  the observed one. Demarque et al. showed that  a  
 moderate  amount  of  convective  core  overshoot (namely $\lambda=0.23 
 H_p$) make  longer the isochrones path, but they are forced to use an age of 
 2.4 Gyr.

We  argue that the isochrones provide us just the locus "permitted" to    
the single stars  in  the  CMD. In order to understand if and how this permitted locus 
is really populated or not (and at what extent) a comparison between  
observed and synthetic CMD is absolutely required. When a suitable mass function 
as well as  binary  
stars  are considered, a very good reproduction of the observed 
sequences of NGC 2420 is obtained (see figure 23). 
In particular the contribution of the binaries
leads to a larger spread in the main sequence. 
Note the effect on the turnoff region: the   
convective path of single stars seems prolonged by the presence of the binary main 
sequence and the bluer region after 
the overall contraction gap is depopulated.
Also in this case we obtain 
an age of 1.6 Gyr which is in very good agreement with the value 
derived from the WD luminosity function cutoff. 

\placefigure{f23int}

\section{CONCLUSIONS}

In this paper we have illustrated the main properties of our latest set of stellar models for intermediate
mass stars as obtained by means
of the FRANEC code (Chieffi \& Straniero 1989). By comparing the most recent evolutionary sequences computed by 
using different evolutionary codes and/or input physics,
we found a rather large disagreement, which is partially due to the influence of
the theoretical assumptions on the 
estimated extension of the convective regions. We would again
remark that not only the difference in the adopted convective algorithm
(Schwarzschild criterion, overshooting, semiconvection and the like) is responsible of
such a disagreement. The connection among the various ingredients of the model cooking must be understood in order
to recognize the origin of the theoretical uncertainties.
In some case we found that models obtained by using very different schemes for the treatment of the
convective instabilities are more similar than models obtained by using the same algorithm, but different input 
physics (eos, opacity, nuclear reaction rate and the like).

In spite of this disagreement, since the brighter is a model the lower is the lifetime,
many differencies are smoothed away when transposing the evolutionary tracks into the 
isochrones. This has been already shown in the previous section where we compare 
our theoretical isochrones and those by Bertelli et al (1994) with the Color Magnitude diagrams of some
well studied Open Clusters (see figure 21). Although the evolutionary features of these two sets of models are rather different,
the resulting isochrone fittings are quite similar in the two cases.  
This means that a quite similar isochrone path 
may be obtained simply by rescaling the mass (or the age). Such an occurrence is clearly illustrated in 
the example reported in figure 24. In this figure we compare our isochrones of 0.7 Gyr with the ones of Bertelli
et all having 0.8 Gyr.

\placefigure{f24int}

We recall that the Bertelli isochrones were obtained by assuming a moderate amount of overshooting, whereas our 
models do not include any extra mixing with respect to the instability boundary. As discussed in the
previous section, the only evident difference is in the shape of the turnoff. 

Another quantities which is well established in the framework of the current theory of the stellar evolution is the 
core mass attained at the beginning of the AGB phase. In fact, since the luminosity of an off main sequence star
is mainly controlled by the size of its He-core mass, the H-burning shell have more time to advance in mass in 
models with lower core mass,

Let us finally comment that the use of the Color Manitude diagrams to check the reliability of a 
particular set of models cannot be barely made by means of the isochrone fitting. In fact the best photometric
studies of open clusters include few thousand of stars. Then the "permitted" locus do not necessarily coincide with the
"populated" locus. In addition many Open Clusters have a huge population of binary stars which contribute to determine the 
shape of the observed Color Magnitude diagrams. The case of NGC2420, as discussed in the previous section, is a template
of such a situation.

\newpage

\centerline{\bf FIGURE CAPTIONS}
\vspace{1cm}

\figcaption[f1.eps]{Evolutionary tracks for Z=0.02 Y=0.28
\label{f1int}}

\figcaption[f2.eps]{Evolutionary tracks for Z=0.006 Y=0.26 
\label{f2int}}

\figcaption[f3.eps]{Evolutionary tracks for Z=0.001 Y=0.23 
\label{f3int}}

\figcaption[f4.eps]{Evolutionary tracks for Z=0.0001 Y=0.23 
\label{f4int}}

\figcaption[f5.eps]{The evolution of the central temperature 
versus the central density for Z=0.02 and Y=0.28.
\label{f5int}}

\figcaption[f6.eps]{The evolution of the central temperature 
versus the central density for Z=0.006 and Y=0.26.
\label{f6int}}

\figcaption[f7.eps]{The evolution of the central temperature 
versus the central density for Z=0.001 and Y=0.23.
\label{f7int}}

\figcaption[f8.eps]{The evolution of the central temperature 
versus the central density for Z=0.0001 and Y=0.23.
\label{f8int}}

\figcaption[f9.eps]{The evolution of the convective regions (hashed areas) 
for the 2.5 ${M_\odot}$ of solar chemical composition.
\label{f9int}}

\figcaption[f10.eps]{The evolution of the convective regions (hashed areas) 
for the 4 ${M_\odot}$ of solar chemical composition.
\label{f10int}}

\figcaption[f11.eps]{The evolution of the convective regions (hashed areas) 
for the 7 ${M_\odot}$ of solar chemical composition.
\label{f11int}}

\figcaption[f12.eps]{The evolution of the convective regions (hashed areas) 
for the 9 ${M_\odot}$ of solar chemical composition.
\label{f12int}}

\figcaption[f13.eps]{The He core mass at the He ignition as a function
of the total mass. 
\label{f13int}}

\figcaption[f14.eps]{He-burning lifetime versus the stellar mass for Z=0.02 and Y=0.28.
\label{f14int}}

\figcaption[f15.eps]{The He core mass at the beginning of the 
TP-AGB phase as a function of the total mass. 
\label{f15int}}

\figcaption[f16int.eps]{The final masses: the He core masses at the beginning of the 
TP-AGB phase (squares); the residual massses at the end of the AGB (triangles); initial/final mass
relation by
Weideman (1987) for the galactic disk (dotted line) and for the Magellanic Clouds (dashed line);
the initial/final mass relation updated by Herwig (1995) for the galactic disk (solid line).  
The numeric labels indicate the number of thermal pulses occurring before the
end of the AGB phase.
\label{f16int}} 

\figcaption[f17.eps]{Comparison among the present evolutionary tracks and those presented by
Castellani, Chieffi \& Straniero (1990 and 1992).
\label{f17int}}

\figcaption[f18.eps]{Comparison among the present H-burning lifetimes (pp) and the those
by Bressan et al (1993, B93) and Shaller et al. (1992, S92). The differencies (in percent)
for each masses are reported. 
\label{f18int}}

\figcaption[f19.eps]{As in figure 18, but for the He-burning lifetimes.
\label{f19int}}

\figcaption[f20.eps]{Comparison among various theoretical core masses at the end of the E-AGB:
present paper (solid line), Lattanzio (1991) Y=0.20 (squares), Lattanzio (1991) Y=0.30 (triangles),
Bressan et al. (1993) (circles). 
\label{f20int}}

\figcaption[f21.eps]{Isochrones fitting to the Pleiades by using our isochrones
(left panel) and those of Bertelli et al. (1994, right panel). 
\label{f21int}}

\figcaption[f22.eps]{Isochrones fitting to NGC2420.
\label{f22int}}

\figcaption[f23.eps]{True and synthetic CMDs. Upper-left panel: 
the observed CMD of NGC2420. Other three panels: synthetic CMDs as obtained 
under different assumption about the exponent (${\alpha}$) of the mass 
function. In computing these synthetic diagrams we have assumed a 30$\%$ 
of binary stars and an age of 1.6 Gyr.  
\label{f23int}}

\figcaption[f24.eps]{Comparison between our isochrones for an age of 0.7 Gyr (no overshoot) 
and the one by Bertelli et al. (1994) for 0.8 Gyr (moderate overshoot). Both these isochrones have 
Z=0.02 and Y=0.28.
\label{f24int}}

\end{document}